\documentclass[aaspp4]{aastex}

\newcommand{\om}{$\omega$~Centauri }
\newcommand{\oc}{$O-C$ }

\shorttitle{Period changes in \om}
\shortauthors{Jurcsik et al.}

\begin{document}

\title{Period changes in \om  RR Lyrae stars}

\author{J. Jurcsik\altaffilmark{1}, 
C. Clement\altaffilmark{2}, E. H. Geyer\altaffilmark{3} and I. Domsa\altaffilmark{1}}

\altaffiltext{1}{Konkoly Observatory of the Hungarian Academy of Sciences, 
1525, Budapest, PO Box 67, Hungary \email{jurcsik@konkoly.hu; domsa@konkoly.hu}}
\altaffiltext{2}{Department of Astronomy, University of Toronto, Ontario, 
M5S 3H8, Canada \email{cclement@utoronto.ca}}
\altaffiltext{3}{Hoher List Observatory, University Bonn, D-54550 Daun, Germany}

\begin{abstract}
A century of observations of \om allows us to accurately follow the 
period changes of a large sample of variable stars. Although the period 
changes are composed of monotone and irregular changes, period increases 
dominate in most of the RRab stars. The typical period increase rates are 
of the same order of magnitude as model calculations predict for the evolved, 
redward phase of horizontal branch evolution. For the 44 well observed RRab 
stars which show monotonic period changes, a mean rate of 0.15 d/Myr has 
been found. The period changes of the first overtone RRc type stars show a 
much more complex, irregular behavior.
\end{abstract}


\newpage
\section{Introduction}

One of the principal goals of globular cluster photometric studies carried 
out during the past century was to determine period changes of RR Lyrae stars. 
The hope to reveal evolutionary changes motivated many extensive works. 
(For details see e.g. \citet{sm95} and \citet{rs97}.) However, attempts to 
connect the observed period changes directly with the different phases of 
horizontal branch (HB) evolution have not led to indisputable conclusions. 
Neither the large period decreases nor the random and/or abrupt period changes 
that have been detected in many clusters can be explained on the basis of steady 
evolutionary effects. 

The basic tool in the determination of period changes of variable stars
is the \oc diagram. This plots the difference between the observed times ($O$)
of a particular phase (e.g. maximum) of the light-curve and the predicted
times of the same phase ($C$) calculated according to an accepted ephemeris.
A nonlinear \oc indicates period changes while a linear \oc with a non-zero 
slope points to inaccuracy of the adopted period.  

\citet{ste34} showed that period-noise can have a 
cumulative effect in the \oc of variable stars, thus mimicking true period 
changes. However, in case of RR Lyrae stars \citet{bdd65} concluded that their 
\oc diagrams cannot be wholly explained as the result of period-noise.  
Nevertheless, \citet{lee91} showed that the mean period change rates observed 
in globular clusters were consistent with synthetic HB models if the HB types 
of the clusters were also considered. However, accurate determinations of 
period changes are available for only a small sample of clusters and for many 
of these, the derived period change rates vary from one study to another 
(see e.~g. the $0.07$ d/Myr difference in the mean period change rate for 
NGC~7006 given  by \citet{we92} and \citet{we99}). Thus it is still an 
important task to define accurate period change rates in globular clusters 
by using not only the most recent data, but also any earlier observations, 
even if they can only be reconstructed from published light curves or epochs 
of maximum light. Searching for unpublished cluster observations and making 
them accessible for future studies is very worthwhile.

The preference for period increases in \om was first noticed by \citet{ma38}.
\citet{be64} strengthened this result and raised the possibility of an 
evolutionary explanation. The recent CCD observations of the variables 
\citep{kal97} within the framework of the OGLE (Optical Gravitational Lensing 
Experiment) project have lengthened the time-base of the observations to more 
than one hundred years, thus making it possible to extend the earlier period 
change studies.

\section{Data}

We used all known and available photometric information on \om variables 
including the original photographic data of \citet{ba02,ma38,di65,di67} and 
the recent CCD V observations of \citet{kal97}. In addition, the scarce 
photoelectric V observations made by \citet{stu75,stu78} and \citet{eg61} of 
V92 proved to be useful in some cases for defining an \oc value at an epoch 
when no other observations were obtained. 

\citet{be64} published \oc values for 47 RRab stars based partly on observations 
obtained with the Yale-Columbia 26-inch refractor in 1947, 1948, 1956, 1958, 
1961, 1962 and 1963. The individual magnitudes were not included in her paper. 
However, we have obtained most of the unpublished data for these 47 RRab stars 
and for a further 41 variables, mainly RRc stars. In addition, magnitudes 
derived from observations made with the same telescope in 1965 and 1967 have 
been obtained. In Table~\ref{belserene}, a sample of the magnitudes for the 
first 11 stars are listed; the rest are available in the electronic edition. 
Observations of some variables not included in the present study are also 
given in Table~\ref{belserene} (SX Phe: V65, red variable: V164, 
and variables with not enough data to investigate period changes: 
V167 and V169). We used \oc values that \citet{be64} derived from the 
$1922-6$ observations \citep{w40} and for the Yale-Columbia 1961 and 1962 
observing seasons because these original data could not be located. 

\citet{weh82} examined the period changes of BL~Her type stars in globular 
clusters. To give the most complete sample of the period changes, we have 
also included the \om stars they studied in our present work, even though 
new data have not been obtained for all of the variables.

About 400 photographic B plates were taken by C. Clement with the University 
of Toronto's 61 cm (f/15) Helen Sawyer Hogg telescope at the Las Campanas 
Observatory of the Carnegie Institution of Washington from 1972 to 1986. 
The plates were measured on a Cuffey iris astrophotometer. As these 
observations span more than a decade, they have crucial importance for 
following the period changes in the second half of the century. 
In Table~\ref{lacampanas}, part of these data for the first 14 variables 
are given. The entire data set is presented in the electronic edition.   

Photographic B and V observations obtained by E. H. Geyer with the 
80/90/300cm ADH Baker Schmidt camera of the Boyden Observatory in the 1962 
observational season \citep{gsz70} were also used. The photometric data of 
these observations have not been published, but recovery of the original 
measurements is possible for 18 variables out of which 16 are used in this
work. The other two stars are V65 (SX Phe), and V167. 
Table~\ref{adh} lists these photographic B data. For most of the observed 
stars only folded light-curves are available. These folded curves could 
be used to determine phase relation only if the assumed period and initial 
epoch were known without doubt. 

In Table~\ref{data-all} the different observations of the \om variables are 
summarized. The new variables discovered  by \citet{kal97} are not included 
in Table~\ref{data-all} because there is no previous record for determining 
the period variation. RR Lyraes with not enough data to construct any reliable 
segment of their \oc are also omitted (V80,110,114,135,136,137,141,143,145,146,147,153,154,
156,157,158,158,169 and all the variables after V170).

\section{Phase shift diagrams, period changes and folded light-curves}

We used the original photometric data to determine phase shifts whenever 
they were available. (These sources are denoted by X in Table~\ref{data-all}.) 
The CCD V \citep{kal97} observations were used to define the normal curves 
and the phase shifts for the other epochs were measured relative to these 
curves. Both vertical and horizontal shifts were derived for all the other 
photometric data in order to match the normal curve best. The most deviant 
magnitudes were omitted from each dataset. We applied a single vertical 
(zero-point) shift to the different observations, with the exception of 
the OGLE measurements. When the same star appeared in different OGLE fields, 
these data were treated separately and differences in their zero points were 
also determined. 

For determining \oc values (horizontal shifts) the datasets were divided 
into shorter segments ($1-3$ years long). (For simplicity, we denote \oc 
as the phase shift between the light-curves as a whole instead of the 
traditional definition of reading phase shifts only between special parts 
of the light-curves.)  If there were no OGLE observations of the star, 
the phase shifts were measured relative to Martin's (1938) photographic 
$m_{pg}$ curves.
 
As phase shifts between photographic B,V, photoelectric and CCD V observations 
were measured, it was also important to check the effect of data inhomogenity 
on the results. By determining the optimal horizontal and vertical shifts 
between the B and V light-curves of the recent CCD B and V observations of 
the M2 variables \citep{lc99}, a mean phase shift of 0.0021 d (with 0.0039 rms) 
between the V and B light-curves of the 19 RRab stars was found. The shift 
for the 12 RRc stars is 0.0001 d (0.0021 rms). Consequently, although the 
times of maxima systematically deviate in B and V, when the phase shifts of 
the whole light-curves are considered then no significant difference according 
to the colors occur. The \oc ambiguity arising from data inhomogeneity is 
therefore less than $1-2\%$ of the period.

\oc values were derived from the data compiled in Table~\ref{data-all} at all
of the epochs for which it was possible to obtain a reliable value. 
Observations of \citet{ba02} and \citet{ma38} were divided into two parts, 
while the Las Campanas observations allowed for the determination of six 
independent \oc points in most cases. Observations from the different OGLE 
fields were used independently. If the observations of one of the OGLE 
fields were extensive enough, \oc-s were determined for two of its segments. 
The \oc values given by \citet{be64} for $1922-6$, 1961 and 1962
(JD~2423530, 2437454 and 2437727, respectively) were appropriately transformed 
if the \oc curve was constructed with a period different from the one she used. 
Similar transformations were applied for the ADH Schmidt data in cases where 
only the folded light-curves were available. 

We constructed the \oc curve by using an estimated value of the average 
period over the century ($P_a$) in order to obtain a symmetrical form and 
fitted the \oc points by different order polynomials. The \oc can be 
expressed as
\begin{equation}
O-C=\int P(E) dE - E P_{a} ={1\over P_{a} }\int P(t) dt - t
\end{equation}
where $E$ is the epoch number, and $P_a$ is the period used to determine \oc.

If we approximate the \oc with a polynomial in the form of
\begin{equation}
O-C=\sum_{i=1}^{k} c_{i} t^{i-1},
\end{equation}
then the temporal period can be determined using the $c_i$ coefficients:  
\begin{equation}
P(t)=P_{a}\sum_{i=2}^{k} (i-1) c_{i} t^{i-2} +P_{a}.
\end{equation}
If it is possible to determine the period with sufficient accuracy 
at the date of the observations, we also have a direct measure of the 
period changes for comparing with the above fit. 

Furthermore, if we transform the time according to
\begin{equation}
t'=t-\sum_{i=1}^{k} c_{i} t^{i-1}, 
\end{equation}
then all the data can be folded with a single period because,
from Eq. 2, it follows that the \oc term vanishes. 

Thus we have two methods for checking the validity of both the \oc 
values and their trial polynomial fits, namely 
(a) a comparison of the direct period determinations with the temporal 
periods predicted according to the derivative of the \oc and 
(b) the scatter of the folded curve of the time transformed data.
This is essential when an ambiguity in the cycle number counts occurs.

Figs. 1a-l, 2a-h and 3 show the results for RRab, RRc and RR Lyrae-like stars
(which were formerly called BL~Her type variables), respectively. In the left 
panels which show the \oc data and their polynomial fits, the filled circles 
denote \oc points derived from the phase shifts of the original data while open 
circles are for Belserenes's (1964) \oc values for $1922-6$, 1961 and 1962 
and for phase shifts which were read from Geyer's folded light-curves. 
If the above checks could not help in deciding cycle numbers, and/or there 
remained significant uncertainties in the \oc curve or in its polynomial fit, 
the fit is shown by a dashed line. The \oc were divided into two overlapping 
parts which were fitted separately with different order polynomials in the 
cases of some RRc stars showing very complex \oc (V10, 64, 76 ,95, 126).
This procedure avoids strong discontinuities in the \oc and period fits. 
However, the possibility that the period changes are in reality abrupt
cannot be excluded. 

In the middle panels, period changes calculated from the \oc fit according to 
Eq. 3 (dashed lines) are shown and compared with the results of direct period 
determinations. Whenever the observations span over two or more consecutive 
seasons, then direct period determinations with sufficient accuracy may also 
become possible. In some cases periods were determined from single observing 
seasons, but usually the accuracy of these period determinations was not good 
enough to make any use of them. When no ambiguity in the \oc values has arisen, 
i.e. $P_a$ seemed to be correct, this period was taken as an initial trial 
period for each data set. Periods in the vicinity of $P_a$ were checked and 
the one that gave the light-curve with the smallest $rms$ scatter in a least 
squares solution was adopted.
In those cases when $P_a$ might be questioned, i.e. cycle number 
ambiguity has arisen indicating that aliasing might seriously affect the
period determinations, a period search using the MUFRAN package \citep{kol90}
was performed. If there was more than one possible peak in the amplitude 
spectra of the Discrete Fourier Transform (DFT), the peak which was closest 
to the period determined from the other observations was chosen if it was 
consistent with the \oc fit as well.

The error range of the periods was determined by checking the residuals ($rms$) 
of the folded curves using periods close to the one accepted according to
the procedure described above. The upper and lower limits were set to the 
periods for which the scatter of the folded curves increased by $10\%$. 
For the best light-curves (some of Martin's and Kaluzny's observations if 
more than 200 data-points were available, and the quality of the light-curve 
was high enough to fit a higher order ($>7$) Fourier sum), only $5\%$ 
was allowed. 
These percentages can be justified as the $90\%$ confidence region of the 
period assuming normal errors, correspond e.g. to $5\%$, $8\%$ and 
$10\%$ increases in the $rms$ scatter of the fit in cases of N=250, M=15; 
N=120, M=10, and N=60, M=5 (N:number of data, M: order of the Fourier fit), 
respectively (see details in \citet{numre92}). 

Checking whether the periods determined from Eq. 3 matched the actually 
observed periods within their error ranges could help in many cases 
to decide the proper cycle numbers.

The folded curves of all the photometric data derived from the time-transformation  
of Eq. 4 are shown in the right panels of Figs. 1-3. As a result of this 
transformation all the data could be brought into phase with the normal curve. 
In those cases where the cycle number or the shape of the \oc fit was dubious, 
the solution that was accepted was the one that resulted in smaller scatter on 
the folded curve of the time-transformed data. We recall that these folded 
curves were matched to CCD V or to photographic $m_{pg}$ normal curves depending 
on whether the star was measured by the OGLE project or not.

The scatter of the time transformed data and the agreement between 
the observed periods and the derivative of the \oc fit  served as 
checks on the correctness of the \oc and its polynomial fit. 
Different order polynomials with different weightings, interpolations 
and extrapolations were tested in order to get the most reliable solution. 
The aim was to determine a more precise \oc curve which also enabled us to 
estimate the contribution of random changes to the global period changing term.
 
In Table~\ref{oc-dat} the \oc values, the order of the \oc fit and the actual 
periods measured and their possible ranges at different epochs are given for V3. 
The results for all of the other variables studied are available in the 
electronic edition which presents Table~\ref{oc-dat} in its complete form. 

The period change rates defined as $\beta=<{\Delta P\over\Delta t}>$ are 
determined by using the polynomial fits to the \oc curves. Temporal 
periods at $\Delta t=1000$~d intervals are calculated according to Eq. 3 
and the mean of the resultant ${\Delta P\over\Delta t}$ values is taken 
as $\beta$.

The period change rates defined as $\beta=<{\Delta P\over\Delta t}>$ are 
calculated by using the polynomial fits to the \oc curves. From the temporal 
periods obtained according to Eq. 3, ${\Delta P\over\Delta t}$ ($\Delta t=1000$~d) 
values are determined and their mean is taken as $\beta$. Measuring $\beta$ in
this way gives practically the same result as calculating $2 c_3P_a$ from a 
quadratic \oc fit where $c_3$ is the coefficient of the second order term or 
from measuring the period difference  between the first and last observed epochs. 

It is difficult to determine whether a period change is actually random 
or monotone. However, the $rms$ scatter of the calculated 
${\Delta P\over\Delta t}$ values during the century 
($\sigma_{\Delta P\over\Delta t}$) may give some indication.
If the scatter is large, the changes are probably random.
In the discussion that follows, we assume that no significantly 
monotonic period change occurs if
\begin{equation}
|<{\Delta P\over\Delta t}>| \le 2 \sigma_{\Delta P\over\Delta t}
\,\,\, i.e. \,\,\,
\sigma_{\Delta P\over\Delta t}/|\beta| \ge 0.5.
\end{equation}

$\beta$ values (in $10^{-10}$ days/day and in days/Myr units), 
the normalized period variations $\alpha= \beta /P_a$ in $10^{-10}$/day, 
$\sigma_{\Delta P\over\Delta t}$, and the 
$\sigma_{\Delta P\over\Delta t}/|\beta|$ ratio for all the variables 
studied are given in Table~\ref{beta}. A comparison of these values with 
Figs. 1-3 demonstrates that our choice of selecting monotone period changes 
according to the Eq. 5 criterion gives reasonable results. 

An examination of the light-curves of all the variables in the course of the 
above procedure revealed that Blazhko behavior (amplitude and/or phase modulation) 
is manifested in many cases. Stars showing definite modulations are marked in 
Figs. 1-3 and are encoded in column 3 of Table~\ref{beta}. Column 4 of 
Table~\ref{beta} indicates ambiguous \oc fits, while column 5 refers to the 
metallicity according to Jurcsik (1998b, see also Table~\ref{paramab}).

\subsection{RRab stars}\label{ab}

The present investigation confirms Belserene's (1964) result that most of 
the \om RRab stars have steadily increasing periods. This is illustrated 
in Fig.~\ref{fig4} where we show a histogram of the $\beta$ distribution for the 
stars that have a well defined \oc fit with monotonic changes exceeding 
the period noise. Of the 71 RRab stars listed in Table~\ref{data-all} and 
Table~\ref{oc-dat}, 27 were excluded from the plot. Four (V9,11,56,59) have 
ambiguous \oc curves (all of them exhibit Blazhko behavior) while for eight 
others (V91,106,132,134,139,144,149,150), there are not enough data to draw 
any firm conclusion about their long term period changes. In addition,
those stars that did not show significantly monotonic changes in their 
periods according to the criterion of Eq. 5 are also omitted. V56 and V84 
were also excluded because their radial velocities indicate that they are 
non-members \citep{li81,le00}.

Altogether, there remain 45 RRab stars for which monotonic period change 
rates seem to have been followed. Nine have decreasing periods; one of them 
(V104) has a steady period decrease at a very high rate 
($5.4\times 10^{-8}$ days/day). The period of this star, however, is unusually 
long for an ordinary RRab star ($0.87$ d). Thus its inclusion in the 
sample may be questioned. The other eight stars have periods decreasing 
at $10^{-10}$ and $- 10^{-9}$ days/day magnitude rates. Period increases at 
similar rates have been detected in 30 of the 45 stars. For the other six,
no change or very slight positive period changes were detected 
($0.0 \le \beta \le 10^{-10}$ days/day). The mean $\beta$ of the 44 stars 
(omitting V104) is $4.1\times 10^{-10}$ days/day (0.15 days/Myr), the median 
is 3.8 or 0.14 in the respective units.

It was shown by \citet{j98b}, that in contrast to the large chemical 
inhomogeneity of the \om stars, the majority of the RRab stars have 
surprisingly homogeneous metallicity. Using the empirical formula that 
relates [Fe/H] to Fourier coefficients \citep{jk96}, metallicities of 48 
RRab stars observed by \citet{kal97} were derived. As no data were given 
in \citet{j98b} we list the physical parameters of RRab stars calculated 
from the light-curves \citep{j98a,kj97,kj96,jk96} in Table~\ref{paramab} 
(see also Sect.~4). In this sample, 40 stars have similar metallicities;
their [Fe/H]$ = -1.54\pm0.08$~dex. Of these 40, reliable \oc could be 
determined for 39 stars. Considering the period changes of only these 
variables, the tendency for the periods to increase is also clear. 
Positive $\beta$ values have been detected in 24 stars, only 4 have 
negative ones, 1 does not show any period change and irregular period 
changes are observed in 10 cases. There was one star (V99) found to be 
more metal poor; its steady period increase rate is 
$\beta=46.0\times10^{-10}$~days/day. Of the 7 relatively metal rich stars 
excluding the non-member V84 and the peculiar V104, 3 have shown steady period 
increases while only one is with decreasing period and the period change of 
one star is not monotonic.

Regarding the Blazhko type behavior, we have found that 14 out of the 
71 RRab stars clearly exhibit some type of modulation in their light-curves. 
Excluding the non-member V56, there is no sign of monotonic period change 
in five of the stars, while out of the eight variables showing monotonic 
period variation, a decreasing period has been observed for only one (V120). 
Although the periods of the \om RRab stars span a very wide range 
($0.47-0.89$ d), all the Blazhko variables have periods less than 0.7 d.

In Fig.~\ref{fig5} the distribution of the normalized period changes ($\alpha$)
is plotted as a function of the period. Stars without a definite sign for their 
period changes are  arbitrarily placed at $\alpha=-70$. These stars are also 
found only among the shorter period variables, most of them have periods within 
a very narrow ($0.59-0.67$ d) period range. This unusual coupling of the periods 
of the Blazhko variables with the periods of stars showing random period variations 
might probably indicate some kind of physical relation between these phenomena.

In conclusion, we have found that, although random variations dominate the 
period changes of 14 stars, period increases dominate in most of the remaining 
stars. Among the RRab variables exhibiting steady period changes, only 9 have 
decreasing periods while 34 have increased their periods during the past century.

\subsection{RRc stars}

We have constructed \oc plots for 48 RRc type variables, but out of which
26 must be taken with caution because of cycle number ambiguity or not enough 
time coverage. The RRc sample shown in Figs. 2a-h clearly indicates that the 
period changes of RRc stars are more complex than those of the fundamental 
mode (RRab) stars. Out of the RRc stars with well defined \oc curves, period 
increases occur in 10, period decreases in 4, random period changes in 7 cases, 
and 1 star does not show any period change. The equivalent numbers for the 26
stars with less certain \oc~-s are 4, 3, 17, and 1, respectively, if V168 is not 
included because it is not a radial velocity member. 

A thorough examination of the light-curves of the RRc stars revealed the existence 
of some kind of modulation in at least 8 out of the 48 first overtone variables. 
This is a similar percentage ($17\%$) to the occurrence of Blazhko behavior in 
the case of RRab stars ($20\%$).

We call attention to a special group of variables V47, V68, and V123 
(showing also light-curve variation) which we included in the RRc group because 
of their amplitudes and light-curves shapes, although it was shown in \citet{cr00} 
that their $\varphi_{21}$ Fourier parameters are anomalously large. 
\citet{gsz70} also noticed that some of the RRc stars define a bluer ridge in the 
B$-$V.$vs$.log$P$ plane, and among others V47 and V123 were found to belong to 
this group. 

The periods of these three stars are unusually long ($0.47-0.53$ d), 
which would rather indicate fundamental mode oscillation and their mean magnitudes 
are among the brightest of the RRc sample. V47 and V68 are 0.1 and 0.2 mag brighter 
than any of the others and the mean magnitude of V123 corresponds to that of the 
brightest among the other variables. Though the period changes of these stars are 
irregular according their $\sigma_{\Delta P\over\Delta t}/|\beta|$ parameter, in 
V68 and V123 significant period increases (1.9 and 5.5 d/Myr, respectively) have 
occurred during the past century. Thus we may be seeing an indication of their 
more evolved status, although this is not as marked as in the case of the RR 
Lyrae-like stars discussed in the next section ($3.3$). These latter stars are 
typically more than 0.5 mag brighter than the RR Lyrae stars. There is, however 
a discernible similarity between the light-curves of these stars and that of 
the faintest evolved star (V92) even though it has a much longer (1.34 d) 
period. At present we can safely conclude that further studies, especially 
accurate color determinations are needed in order to clarify the true nature of 
these peculiar objects.

Omitting these 3 variables and stars with too short \oc segments, a similar plot 
of the $\alpha$ distribution of the RRc variables as seen in Fig.~\ref{fig5} 
for the RRab stars is shown in Fig.~\ref{fig6}. Again, stars with irregular 
period changes tend to be found among the longer period RRc stars where the 
modulated (Blazhko) variables are.

\subsection{RR Lyrae-like stars}

\citet{weh82} made a thorough analysis of period changes of BL~Her stars in 
globular clusters. They classified  5 of the \om variables as BL~Her type 
and found period increase rates in the range of $10^{-9}-10^{-8}$~days/day  
for all of them. 

The name BL~Her is no longer used for these longer period RR Lyrae-like
stars (short period population~II Cepheids), but the classification scheme 
has not yet been settled (see e.g. \citet{sdt94,nnl94}). In this category, 
we consider the AHB1 stars (according to the nomenclature of \citet{gs96}). 
AHB1 stars are `above horizontal branch' stars which are either significantly 
brighter ($0.5- 1$~mag) or have longer periods than the cluster RR Lyrae stars. 
Thus we also add V52 to this group because it is $0.5$ mag brighter than other 
RRab stars with similar period (0.66~d). Based on its period we may also include 
V104 (P=0.87~d, $\beta=-5.4\times10^{-8}$~days/day) into this group, however 
its very large monotonic period decrease rate represents a disturbing extreme 
in any of the groups.
 
We also included V43 and V48 in our study, although no new observations were
available. The period change rates we have determined for these stars agree 
within the errors to those of \citet{weh82}, showing the compatibility of our 
method with other period changes studies. The addition of new \citep{kal97} 
data to the earlier measurements strengthens the dominance of the monotonic period 
increases for V60,V61 and V92. Our solutions for V60 and V92 are, however, 
somewhat different from the earlier results \citep{weh82} mainly because we 
also take direct period determinations into account and this yields even 
larger period increase rates.

Thus the period increases observed in these stars are in agreement with their 
evolved HB status. In the last phase of the HB, during evolution 
towards the asymptotic giant branch, large period increases are expected.

\section{Conclusion: Comparison with evolutionary model predictions}

We have shown in Sect.~\ref{ab} that most of the RRab stars have increased 
their periods by some $10^{-11}-10^{-9}$ days/day rate during the past 
century. This is in agreement with evolutionary model predictions for the 
late, redward phase of the horizontal branch evolution. The period changes 
along the post zero age horizontal branch, as derived from the \citet{do92} 
models and from the linear pulsation period formulae of \citet{kb94}, range 
typically from $- 10^{-10}$ to $+ 10^{-9}$ days/day. Within the instability 
strip there is no metallicity and mass value for which the evolutionary 
period decrease rate is larger than some $ 10^{-11}$ days/day, but period 
increases of the order of $10^{-9}$ days/day do occur in the redward phase 
of the blue loops in the HB evolution, especially for the lower mass stars.

The physical parameters derived from the light-curves (Jurcsik 1998b, and 
Table~\ref{paramab}) indicate that most of the RRab stars in \om belong to 
a very homogeneous population, and are already in the evolved (redward) phase 
of their HB evolution. Thus, their comparison with model predictions of unique 
chemical composition models give adequate results, in spite of the fact that 
large chemical inhomogeneities have been detected in \om stars.

In Fig.~\ref{fig7} horizontal branch evolutionary tracks are shown for oxygen 
enhanced models with [Fe/H]$=-1.48$ dex \citep{do92}. Tracks for four 
different masses ($0.66, 0.64, 0.62, 0.60 M_{\odot}$) are shown. The position 
of the 40 RRab stars which were found to have [Fe/H]$=-1.54\pm0.08$ dex 
\citep{j98b} are also given. The luminosities and temperatures of these 
variables (given in Table~\ref{paramab}) were determined by using the empirical 
relations between the physical parameters and the Fourier coefficients of the 
light curves (see \citet{j98a} and references therein). The empirical $\log L$ 
and $\log T$ scales are shifted by 0.10 and 0.016, respectively, in order to 
reach agreement with evolutionary results \citep{jk99}. This transformation 
yields a mean pulsational mass of $0.64M_{\odot}$ with $0.02M_{\odot}$ rms 
scatter for these 40 variables. 

Linear pulsational periods have been calculated for different locations
along the tracks where the fundamental mode period falls within the $0.4 -1.0$~d 
range. Then, evolutionary period changes have been calculated using these 
periods and the evolutionary time scales of the \citet{do92} models. They are 
given in units of $10^{-10}$ days/day in Fig.~\ref{fig7}. The distribution of 
the observed $\beta$ values shown in Fig.~\ref{fig4} for those RRab stars 
which have well defined monotonic period change rates is in very good 
agreement with the evolutionary $\beta$ values calculated within the region 
occupied by these stars.

Summarizing our results, we have found that, for those \om RRab stars that 
exhibit dominantly monotonic period changes, the period change rates can be 
brought into very good agreement with horizontal branch evolutionary model 
predictions. Thus the observed period changes of these stars may directly 
indicate evolutionary effects. However, irregular period changes and one very 
large period decrease rate (similar to those observed in other globular clusters) 
also occur in \om and no satisfactory explanation for these phenomena has yet 
been given. The much less regular period change behaviour of the first overtone 
RRc stars is also not yet understood.

\acknowledgments
We are grateful to the referee, Emilia Belserene, for useful comments 
and suggestions. We also thank E. Belserene for sending some of her 
unpublished magnitudes. C. Clement would like to thank Peter Hanson and 
Tom Wells for making the iris photometer measurements.  
Fruitful discussions with B. Szeidl are highly appreciated. 
We are also grateful for financial support for this work which 
was partially provided by the Hungarian OTKA grants T-024022 and T-030954 
and by the Natural Sciences and Engineering Research Council of Canada.

\newpage

\notetoeditor{Please edit Figures 1a-l,2a-h,3 in 4 figures/page format}

\figcaption[Jurcsik.fig1a.ps]
{Results for RRab stars. \oc (left panels), directly measured period values 
(center) and folded light-curves of all the photometric data (right), after the 
times of the observations were transformed according to the polynomial fit of 
the \oc curve. 2\,400\,000 has been substructed from the JDs.
Dashed lines in the \oc plots indicate an uncertainty in cycle 
number counts or in the shape of the \oc fit. The filled circles denote \oc
points derived from the original photometric data and the open circles denote 
\oc values adopted, or estimated from folded light curves. The notation 'Bl' 
indicates evidence of light curve modulation. The period fits 
in the center panels correspond to the derivative of the \oc fit and not to 
actual fits to the periods. The filled circles represent direct period 
determinations. The magnitude scale in the right panel is CCD V if the star 
was measured by \citet{kal97} and photographic $m_{pg}$ of \citet{ma38} for 
all the other stars.
\label{fig1a-l}}

\figcaption[Jurcsik.fig2a.ps]
{Results for RRc stars. See details in caption Fig. 1a.
\label{fig2a-h}}

\figcaption[Jurcsik.fig3.ps]
{Results for RR Lyrae like stars. See details in caption Fig. 1a.
\label{fig3}}

\figcaption[Jurcsik.fig4.ps]
{Histogram of period change rates ($\beta$) of the 44 RRab stars that
have unambiguous \oc curves and monotonic period changes dominating over 
irregular period noise. The bulk of these RRab stars have continuously 
increased their periods by some $10^{-10}$ days/day rate during the century 
long time-base of the observations.
\label{fig4}}

\figcaption[Jurcsik.fig5.ps]
{Normalized period change rates ($\alpha$) $vs.$ periods of RRab stars. 
Open circles denote Blazhko variables. V104 ($\alpha=-625\times10^{-10}$/day), 
the non-members (V56, V84), and stars that do not have long enough continuous 
coverage on their \oc curves are not shown. Those variables which have some 
ambiguity in their \oc fits are included because their dominant (random or 
monotonic)  changes can be safely determined. Alpha values of those 
variables which exhibit random period changes without significant monotonic 
contribution are arbitrarily set to $-70$ just to show their distribution 
with period. All of these stars and the ones with decreasing periods belong 
to the shorter period group of the RRab sample, similar to the Blazhko type 
variables.
\label{fig5}}

\figcaption[Jurcsik.fig6.ps]
{The same plot as Fig.~\ref{fig5} for the RRc stars. The anomalously long period 
stars (V47,68,123) and V151,V163 and V168 are not shown. Stars with random period 
changes are arbitrarily set to $\alpha=-150$.
\label{fig6}}

\figcaption[Jurcsik.fig7.ps]
{Horizontal branch evolutionary tracks of the oxygen enhanced, [Fe/H]$=-1.48$, 
$M=0.60, 0.62, 0.64, 0.66 M_\odot$ models within the region of the instability 
strip \citep{do92}. Evolutionary period changes along the tracks ($\beta$ in 
$10^{-10}$ days/day unit) are given. Filled circles denote the 40 variables 
with [Fe/H]$=-1.54\pm0.08$~dex and $M=0.64\pm0.02M_{\odot}$ pulsational mass 
values. Long term period changes for 39 of these stars have been determined: 
period increase with $\beta=6.6$ mean rate for 24 stars, decreasing periods 
for 4 stars, and random period changes for 10 stars. One star does not show 
any period change. The largest period increase rates ($\beta>20.0$) are 
observed in the two most luminous stars. The mean period increase rates for 
the stars with positive $\beta$ values are in very good agreement with the 
evolutionary results.
\label{fig7}}

\clearpage

\pagestyle{empty}
\renewcommand{\arraystretch}{0.95}

\begin{deluxetable}{lrrrrrrrrrrr}
\tablecolumns{15}
\tablewidth{0pt}
\tablenum{1}
\tabletypesize{\scriptsize}
\tablecaption{Yale-Columbia observations of \om variables.
The complete version of this table is in the electronic edition of the Journal.
The printed edition contains only a sample.
\label{belserene}}
\tablehead{
\colhead{J.D.$-2\,400\,000$}
&\colhead{V3} 
&\colhead{V7}
&\colhead{V8}
&\colhead{V10}
&\colhead{V12}
&\colhead{V13}
&\colhead{V14}
&\colhead{V15}
&\colhead{V16}
&\colhead{V18}
&\colhead{V19}}
\startdata
32234.585 &14.32&14.42&14.00&14.86&14.70&14.53&14.50&14.15&15.05&15.36&15.10\\ 
32266.500 &14.18&14.52&14.68&14.98&14.58&14.56&14.99&14.72&14.64&14.48&14.72\\ 
32278.462 &14.57&15.03&     &14.92&14.49&15.22&14.59&14.36&14.98&15.02&14.77\\ 
32295.414 &14.88&     &15.41&14.93&14.48&14.38&14.48&14.22&14.54&15.29&15.24\\ 
32296.410 &15.01&15.10&15.42&14.82&14.96&15.09&     &14.67&     &14.83&15.06\\ 
\enddata
\end{deluxetable}

\begin{deluxetable}{lrrrrrrrrrrrrrr}
\tablecolumns{15}
\tablewidth{0pt}
\tablenum{2}
\tabletypesize{\scriptsize}
\tablecaption{La Campanas observations of \om variables.
The complete version of this table is in the electronic edition of the Journal.
The printed edition contains only a sample.
\label{lacampanas}}
\tablehead{
\colhead{J.D.$-2\,400\,000$}
&\colhead{V3} 
&\colhead{V4}
&\colhead{V5}
&\colhead{V8}
&\colhead{V9}
&\colhead{V10}
&\colhead{V13}
&\colhead{V16}
&\colhead{V18}
&\colhead{V19}
&\colhead{V20}
&\colhead{V22}
&\colhead{V23}
&\colhead{V24}}
\startdata
41446.739 &15.08&15.29&15.32&15.32&14.35&14.96&14.14&14.99&14.79&15.32&14.85&15.01&15.32&14.99\\
41447.555 &15.26&14.34&14.54&14.16&15.34&14.97&14.80&14.41&15.29&14.76&15.17&15.14&14.30&14.66\\
41447.655 &14.38&14.92&15.13&15.00&15.38&14.51&15.07&14.61&15.25&15.28&15.28&14.69&15.02&15.10\\
41448.507 &14.26&15.22&14.26&15.33&15.29&14.46&15.11&14.60&14.30&15.15&14.34&14.41&15.04&14.67\\
41449.475 &14.25&14.48&15.41&15.44&15.10&14.80&14.39&14.96&15.26&15.41&15.20&15.05&15.35&14.92\\
\enddata
\end{deluxetable}

\begin{deluxetable}{lrrrrrrrrr}
\tablecolumns{9}
\tablewidth{0pt}
\tablenum{3}
\tabletypesize{\scriptsize}
\tablecaption{ADH Schmidt observations of \om variables.
The complete version of this table is in the electronic edition of the Journal.
The printed edition contains only a sample.
\label{adh}}
\tablehead{
\colhead{J.D.$-2\,400\,000$}
&\colhead{V4}
&\colhead{V10}
&\colhead{V14}
&\colhead{V18}
&\colhead{V19}
&\colhead{V22}
&\colhead{V24}
&\colhead{V36}
&\colhead{V56}}
\startdata
37792.289  & 14.85  & 14.60  & 14.81  & 15.26  & 14.74  &        &        & 14.99  & 15.27  \\
37792.315  & 14.75  & 14.35  & 14.74  & 15.05  & 14.84  & 14.57  & 14.25  & 14.89  & 15.35  \\
37792.358  & 14.82  & 14.25  & 14.96  &        & 14.45  & 14.31  & 14.32  & 14.99  & 15.01  \\
37793.341  & 13.99  & 14.74  & 14.50  & 15.11  & 14.73  & 14.93  & 14.57  & 14.59  & 15.35  \\
37793.368  & 14.17  & 14.45  & 14.52  & 15.16  & 15.00  & 14.88  & 14.77  & 14.62  & 15.49  \\
\enddata   
\end{deluxetable}

\begin{deluxetable}{rrrllllllllllc}
\tablecolumns{14}
\tablewidth{290pt}
\tablenum{4}
\tabletypesize{\scriptsize}
\tablecaption{Photometric data of RR Lyrae and evolved variables in $\omega$ Centauri.
\label{data-all}}
\tablehead{
\multicolumn{2}{c}{Var.\tablenotemark{a}} &
\colhead{} &
\multicolumn{9}{c}{References\tablenotemark{b}} &
\colhead{} & \colhead{} \\ 
\colhead{SH} &\colhead{K} &\colhead{} &\colhead{1} &\colhead{2} &\colhead{3} 
&\colhead{4}
&\colhead{5} &\colhead{6} &\colhead{7} &\colhead{8} &\colhead{9} &\colhead{}
&\colhead{type}}
\startdata
  3 &  184 && X&X&X&0&      X0&$-$&X&X&X         && ab \\
  4 &   99 && X&X&X&X&     $-$&$-$&X&X&X         && ab \\
  5 &  101 && X&X&X&0&     $-$&$-$&X&X&$-$       && ab \\
  7 &   87 && X&X&X&0&      X0&X&$-$&X&$-$       && ab \\
  8 &  199 && X&X&X&0&      X0&$-$&X&X&$-$       && ab \\
  9 &  183 && X&X&X&$-$&   $-$&$-$&X&X&X         && ab \\
 10 &   98 && X&X&X&X&       X&$-$&X&X&X         && c  \\
 11 &  126 && X&X&$-$&C&   $-$&$-$&$-$&X&X       && ab \\
 12 &  125 && X&X&$-$&$-$&   X&$-$&$-$&X &X      && c  \\
 13 &  188 && X&X&X&0&       X&$-$&X&X&X         && ab \\
 14 &      && X&X&X&X&       X&$-$&$-$&$-$&$-$   && c  \\  
 15 &  124 && X&X&$-$&0&    X0&$-$&$-$&X&X       && ab \\
 16 &  205 && X&X&X&0&       X&$-$&X&X&$-$       && c  \\
 18 &  200 && X&X&X&X&      X0&$-$&X&X&$-$       && ab \\
 19 &  192 && X&X&X&X&       X&$-$&X&X&$-$       && c  \\
 20 &  163 && X&X&X&$-$&     X&$-$&X&X&X         && ab \\
 21 &   97 && X&X&$-$&$-$&   X&$-$&$-$&X&X       && c  \\
 22 &  190 && X&X&X&X&     $-$&$-$&X&X&X         && c  \\
 23 &  146 && X&X&$-$&$-$&  X0&$-$&X&X&X         && ab \\
 24 &  189 && X&X&X&X&       X&$-$&X&X&$-$       && c  \\
 25 &  113 && X&X&$-$&$-$&  X0&$-$&X&X&$-$       && ab \\
 26 &  120 && X&X&$-$&$-$&   X&$-$&X&X&$-$       && ab \\
 27 &  114 && X&X&$-$&$-$&   X&$-$&X&X&X         && ab \\
 30 &   92 && X&X&$-$&$-$& $-$&$-$&$-$&X&X       && c  \\
 32 &   82 && X&X&$-$&$-$&  X0&$-$&X&X&$-$       && ab \\
 33 &  182 && X&X&X&0&      X0&$-$&$-$&X&$-$     && ab \\
 34 &   90 && X&X&X&0&       X&$-$&X&X&X         && ab \\
 35 &   78 && X&X&X&$-$&     X&$-$&X&X&X         && c  \\
 36 &   86 && X&X&X&X&     $-$&$-$&$-$&X&$-$     && c  \\
 38 &  170 && X&X&X&0&      X0&$-$&$-$&X&$-$     && ab \\
 39 &  197 && X&X&X&0&       X&$-$&X&X&X         && c  \\
 40 &  107 && X&X&$-$&$-$&  X0&$-$&X&X&X         && ab \\
 41 &  129 && X&X&$-$&$-$&   X&$-$&X&X&X         && ab \\
 43 &      && X&X&X&$-$&   $-$&$-$&X&$-$&$-$     && evolved\\
 44 &   88 && X&X&X&0&      X0&$-$&X&X&X         && ab \\
 45 &  179 && X&X&X&0&     $-$&X&X&X&$-$         && ab \\
 46 &  180 && X&X&X&0&       X&X&X&X&$-$         && ab \\
 47 &  185 && X&X&$-$&C&   $-$&$-$&$-$&X&X       && c  \\
 48 &      && X&X&X&$-$&   $-$&$-$&X&$-$&$-$     && evolved\\
 49 &  165 && X&X&X&$-$&     X&$-$&X&X&$-$       && ab \\
 50 &  187 && X&X&X&0&       X&$-$&X&X&$-$       && c  \\
 51 &  147 && X&X&$-$&$-$&   X&$-$&X&X&X         && ab \\
 52 &  116 && X&X&$-$&$-$& $-$&$-$&$-$&X&$-$     && evolved\\
 54 &   74 && X&X&$-$&0&    X0&$-$&X&X&$-$       && ab \\
 55 &      && X&X&$-$&0&    X0&X&$-$&$-$&$-$     && ab \\
 56 &      && X&X&X&X&       X&$-$&X&$-$&$-$     && ab \\
 57 &  206 && X&X&X&0&      X0&$-$&X&X&$-$       && ab \\
 58 &  100 && X&X&$-$&X&     X&$-$&$-$&X&X       && c  \\
 59 &   93 && X&X&$-$&$-$& $-$&$-$&$-$&X&X       && ab \\
 60 &  103 && X&X&X&$-$&   $-$&$-$&X&X&$-$       && evolved \\
 61 &  164 && X&X&X&$-$&   $-$&$-$&X&X&$-$       && evolved \\
 62 &  117 && X&X&$-$&$-$&   X&$-$&$-$&X&X       && ab \\
 63 &      && X&X&X&0&      X0&Y&$-$&$-$&$-$     && ab \\
 64 &      && X&X&X&0&       X&$-$&X&$-$&$-$     && c  \\
 66 &   76 && X&X&$-$&$-$& $-$&$-$&$-$&X&X       && c  \\
 67 &   77 && X&X&$-$&X&   $-$&$-$&X&X&$-$       && ab \\
 68 &   73 && X&X&$-$&X&     X&$-$&$-$&X&$-$     && c  \\
 69 &      && X&X&$-$&0&   $-$&X&$-$&$-$&$-$     && ab \\
 70 &  151 && X&X&X&X&       X&$-$&X&X&X         && c  \\
 71 &  157 && X&X&$-$&$-$& $-$&$-$&$-$&X&X       && c  \\
 72 &      && X&X&X&0&       X&$-$&$-$&$-$&$-$   && c  \\
 73 &      && X&X&$-$&0&   $-$&X&$-$&$-$&$-$     && ab \\
 74 &   85 && X&X&X&0&      X0&$-$&X&X&$-$       && ab \\
 75 &   84 && X&X&$-$&0&     X&$-$&$-$&X&$-$     && c  \\
 76 &   79 && X&X&X&0&       X&$-$&X&X&$-$       && c  \\
 77 &   80 && X&X&$-$&C&   $-$&$-$&$-$&X&X       && c  \\
 79 &  204 && X&X&X&0&       X&$-$&$-$&X&$-$     && ab \\
 81 &  195 && X&X&X&0&       X&$-$&X&X&$-$       && c  \\
 82 &  193 && X&X&X&$-$&     X&$-$&X&X&$-$       && c  \\
 83 &   83 && X&X&X&X&       X&$-$&X&X&$-$       && c  \\
 84 &  174 && X&X&$-$&0&     X&X&$-$&X&$-$       && ab \\
 85 &  176 && X&X&$-$&0&     X&Y&$-$&X&$-$       && ab \\
 86 &  158 && X&X&$-$&$-$&   X&$-$&X&X&X         && ab \\
 87 &  144 && X&X&$-$&$-$& $-$&$-$&$-$&X&X       && c  \\
 88 &  210 && X&X&$-$&$-$&  X0&$-$&$-$&X&$-$     && ab \\
 89 &  143 && X&X&$-$&$-$& $-$&$-$&$-$&X&X       && c  \\
 90 &  141 && X&X&$-$&$-$& $-$&$-$&$-$&X&$-$     && ab \\
 91 &      && X&X&$-$&$-$& $-$&$-$&$-$&$-$&$-$   && ab \\
 92 &   72 && X&X&X&0&     $-$&X&X&X&$-$         && evolved \\
 94 &      && X&X&$-$&0&     X&$-$&$-$&$-$&$-$   && c  \\
 95 &  178 && X&X&X&0&       X&X&X&X&$-$         && c  \\
 96 &      && X&X&$-$&$-$& $-$&$-$&X&$-$&$-$     && ab \\
 97 &  159 && X&X&$-$&$-$&   X&$-$&$-$&X&X       && ab \\
 98 &  160 && X&X&$-$&$-$& $-$&$-$&$-$&X&$-$     && c  \\
 99 &  137 && X&X&$-$&$-$&   X&$-$&$-$&X&X       && ab \\
100 &      && X&X&$-$&$-$&   X&$-$&$-$&$-$&$-$   && ab \\
101 &  194 && X&X&X&$-$&     X&$-$&$-$&X&$-$     && c  \\
102 &  153 && X&X&$-$&$-$&   X&$-$&$-$&X&X       && ab \\
103 &      && X&X&$-$&$-$&   X&$-$&$-$&$-$&$-$   && c  \\
104 &  201 && X&X&X&0&      X0&$-$&X&X&X         && ab \\
105 &  198 && X&X&X&X&       X&$-$&X&X&$-$       && c  \\
106 &      && X&X&$-$&$-$& $-$&$-$&$-$&$-$&$-$   && ab \\
107 &  154 && X&X&$-$&$-$&   X&$-$&$-$&X&X       && ab \\
108 &  211 && X&X&$-$&$-$&  X0&$-$&$-$&X&$-$     && ab \\
109 &  133 && X&X&$-$&$-$&   X&$-$&X&X&$-$       && ab \\
111 &  136 && X&X&$-$&$-$& $-$&$-$&$-$&X&X       && ab \\
112 &  150 && X&X&$-$&$-$& $-$&$-$&$-$&X&$-$     && ab \\
113 &  128 && X&X&$-$&$-$&   X&$-$&X&X&$-$       && ab \\
115 &   89 && X&X&X&$-$&   $-$&$-$&X&X&X         && ab \\
116 &  127 && X&X&$-$&$-$& $-$&$-$&$-$&X&$-$     && ab \\
117 &   94 && X&X&$-$&$-$&   X&$-$&$-$&X&X       && c  \\
118 &  109 && X&X&$-$&$-$& $-$&$-$&$-$&X&$-$     && ab \\
119 &  106 && X&X&$-$&$-$& $-$&$-$&$-$&X&X       && c  \\
120 &  104 && X&X&$-$&$-$& $-$&$-$&$-$&X&$-$     && ab \\
121 &  105 && X&X&$-$&$-$&   X&$-$&$-$&X&$-$     && c  \\
122 &  102 && X&X&$-$&$-$&   X&$-$&X&X&$-$       && ab \\
123 &  169 && X&X&$-$&0&     X&$-$&$-$&X&X       && c  \\
124 &  168 && X&X&$-$&X&     X&$-$&$-$&X&$-$     && c  \\
125 &  166 && X&X&X&0&       X&$-$&$-$&X&$-$     && ab \\
126 &  207 && X&X&$-$&X&     X&$-$&$-$&X&$-$     && c  \\
127 &  175 && X&X&X&0&       X&X&X&X&$-$         && c  \\
128 &   91 && X&X&$-$&$-$&  X0&$-$&$-$&X&$-$     && ab \\
130 &   75 && X&X&$-$&C&   $-$&$-$&$-$&X&$-$     && ab \\
131 &  110 && X&X&$-$&$-$& $-$&$-$&$-$&X&X       && c  \\
132 &      && X&X&$-$&$-$& $-$&$-$&$-$&$-$&$-$   && ab \\
134 &      &&$-$&X&$-$&0&    X&$-$&$-$&$-$&$-$   && ab \\
139 &  118 &&$-$&X&$-$&$-$&$-$&$-$&$-$&X&$-$     && ab \\
144 &  112 &&$-$&X&$-$&$-$&$-$&$-$&$-$&X&$-$     && ab \\
149 &      &&$-$&X&$-$&0&    X&X&$-$&$-$&$-$     && ab \\
150 &      &&$-$&X&$-$&$-$&  X&$-$&$-$&$-$&$-$   && ab \\
151 &      &&$-$&X&$-$&$-$&  X&$-$&$-$&$-$&$-$   && c  \\
155 &  145 &&$-$&X&$-$&$-$&  X&$-$&$-$&X&X       && c  \\
160 &      &&$-$&X&$-$&0&    X&X&$-$&$-$&$-$     && c  \\
163 &      &&$-$&X&$-$&0&    X&$-$&$-$&$-$&$-$   && c  \\
168 &  181 &&$-$&$-$&$-$&$-$&X&$-$&$-$&X&$-$     && c  \\
\enddata
\tablenotetext{a}{ SH refers to \citet{sh73} and \citet{cc00} catalogue numbers, K to the \citet{kal97} IDs.}
\tablenotetext{b}{ X: photometry; Y: photometry too few to define \oc value; 
0: folded light curve or \oc data available; C: only folded light curve available without
parameters allowing to determine \oc.}
\tablerefs{
(1) \citet{ba02}; 
(2) \citet{ma38}; 
(3) \citet{di65} or \citet{di67} for evolved stars; 
(4) \citet{gsz70} data published in this paper and read from their folded light-curves; 
(5) \citet{be64} data published in this paper and \citet{be64} \oc values for JD~2423530, 2437454 and 2437727; 
(6) \citet{stu75,stu78,eg61} for V92; 
(7) C. Clement observations published in this paper or \citet{weh82} for evolved stars; 
(8)-(9) observations of different OGLE
fields \citet{kal97}.}
\end{deluxetable}

\begin{deluxetable}{lrrrrrr}
\tablecolumns{7}
\tablewidth{0pt}
\tablenum{5}
\tabletypesize{\scriptsize}
\tablecaption{\oc and period data of \om variables.
The complete version of this table is in the electronic edition of the Journal.
The printed edition contains only a sample.
\label{oc-dat}}
\tablehead{
\multicolumn{6}{l}{Var. \,\,\, \,\, $P_a (d)$\tablenotemark{a} \,\,\, order of the polynomial fit to the \oc}&\\
\colhead{mean JD\tablenotemark{b}} 
&\multicolumn{1}{r}{\oc (d)}
&\colhead{first JD\tablenotemark{b,}\,\, \tablenotemark{c}} 
&\multicolumn{1}{r}{last JD\tablenotemark{b,}\,\, \tablenotemark{c}} &\colhead{N\tablenotemark{c,}\,\, \tablenotemark{d}} &\colhead{$\Delta$m\tablenotemark{c,}\,\, \tablenotemark{e}} &\multicolumn{1}{r}{Ref.\tablenotemark{f}}\\
\colhead{mean JD\tablenotemark{b}}&\colhead{measured period}&\multicolumn{2}{c}{period range}&&}
\startdata
\sidehead{ \,\, V003 \,\,\,  0.84122900  \,\,\, 6}\\
13153.06  &-0.036 &12234.7 &13411.6 & 43 & 0.88 &1\\
14174.72  &-0.064 &13652.6 &14518.5 & 61 & 0.88 &1\\
23530.00  &-0.383 &    0.0 &    0.0 &  0 & 0.00 &5\\
26451.87  &-0.449 &26406.5 &26836.4 &263 &-0.24 &2\\
27907.03  &-0.453 &27891.4 &27988.0 &135 &-0.24 &2\\
32463.33  &-0.486 &32234.6 &32736.2 & 24 &-0.28 &5\\
35670.10  &-0.503 &35607.0 &35682.0 & 27 &-0.28 &5\\
36269.99  &-0.539 &36259.2 &36292.2 & 13 &-0.28 &5\\
37298.90  &-0.498 &36754.2 &37491.2 & 36 &-0.36 &3\\
37454.00  &-0.481 &    0.0 &    0.0 &  0 & 0.00 &5\\
37811.00  &-0.485 &    0.0 &    0.0 &  0 & 0.00 &4\\
38084.97  &-0.457 &38062.1 &38118.1 & 12 &-0.28 &5\\
38885.84  &-0.466 &38876.1 &38909.0 & 11 &-0.28 &5\\
39629.88  &-0.453 &39563.1 &39684.9 & 28 &-0.28 &5\\
41669.38  &-0.385 &41446.7 &41874.5 & 43 &-0.34 &7\\
42359.68  &-0.355 &42074.8 &42594.5 & 53 &-0.34 &7\\
43520.57  &-0.303 &43273.5 &43699.7 & 73 &-0.34 &7\\
44194.27  &-0.272 &44016.6 &44378.8 & 53 &-0.34 &7\\
44993.61  &-0.237 &44750.6 &45470.5 & 79 &-0.34 &7\\
46244.63  &-0.175 &45786.8 &46562.7 & 54 &-0.34 &7\\
49524.49  &-0.016 &49515.5 &49525.7 & 28 &-0.02 &8\\
49831.88  & 0.000 &49818.5 &49922.5 &201 & 0.00 &9\\
\\
   13752 &0.84120425 &0.84118370 &0.84122730&&&\\
   26945 &0.84122507 &0.84121650 &0.84123400&&&\\
   36544 &0.84122675 &0.84121620 &0.84124020&&&\\
   39105 &0.84125025 &0.84123810 &0.84126290&&&\\
   42685 &0.84126670 &0.84125840 &0.84127480&&&\\
   45129 &0.84126682 &0.84125890 &0.84127450&&&\\
   49794 &0.84127000 &0.84125740 &0.84128260&&&\\
\enddata
\tablenotetext{a}{ The \oc values are calculated according to this period.}
\tablenotetext{b}{ JD$-$2\,400\,000}
\tablenotetext{c}{ 0.0 values are for references when the original photometric data are not available.}
\tablenotetext{d}{ Number of data points used.}
\tablenotetext{e}{ Magnitude-shift applied to vertically match data to the normal curve.}
\tablenotetext{f}{ Reference numbers are the same as in Table~\ref{data-all}.}
\end{deluxetable}

\begin{deluxetable}{rccccrrrrr}
\tablecolumns{10}
\tablewidth{0pt}
\tablenum{6}
\tabletypesize{\scriptsize}
\tablecaption{Period changes parameters of \om variables.\label{beta}}
\tablehead{
\colhead{Var.\tablenotemark{a}}
&\colhead{$P_a (d)$} 
&\colhead{\, Bl\tablenotemark{b}\,\,\,}
&\colhead{$O-C$\tablenotemark{c}}
&\colhead{\hskip -2mm[Fe/H]\tablenotemark{d}}
&\multicolumn{2}{c}{$\beta =<{{\Delta P}\over{\Delta t}}>$} 
&\colhead{$\alpha = \beta /P_a$}
&\colhead{$\sigma_{{\Delta P}\over{\Delta t}}$}
&\colhead{$\sigma_{{\Delta P}\over{\Delta t}}/|\beta|$}\\
&\colhead{mean period}
&
&
& 
&\colhead{$d/10^{10} d$} 
&\colhead{$d/Myr $}
&\colhead{$1/10^{10} d$}
&
&}
\startdata
\sidehead{RRab stars}
  3&   0.8412290& 1&1&1&  21.750&   0.794&  25.855&    1.975&   0.09\\
  4&   0.6273165& 1&1&1&   3.733&   0.136&   5.951&    0.649&   0.17\\
  5&   0.5152828& 0&1& &   1.617&   0.059&   3.137&    0.832&   0.51\\
  7&   0.7130227& 1&1&1&   6.833&   0.249&   9.584&    0.901&   0.13\\
  8&   0.5212930& 1&1&1&  16.950&   0.619&  32.515&    0.696&   0.04\\
  9&   0.5233400& 0&0& &  32.314&   1.179&  61.745&   20.363&   0.63\\
 11&   0.5648060& 0&0& &   5.042&   0.184&   8.926&    0.008&   0.00\\
 13&   0.6690535& 1&1&1&   2.528&   0.092&   3.778&    2.725&   1.08\\
 15&   0.8106220& 1&1&1&  20.406&   0.745&  25.173&    0.004&   0.00\\
 18&   0.6216686& 1&1&1&   1.217&   0.044&   1.957&    1.585&   1.30\\
 20&   0.6155540& 1&1&1&   2.222&   0.081&   3.610&    2.726&   1.23\\
 23&   0.5108675& 1&1&0&   2.122&   0.077&   4.154&    1.822&   0.86\\
 25&   0.5884890& 1&1&1&   8.842&   0.323&  15.024&   14.117&   1.60\\
 26&   0.7847145& 1&1&1&   4.531&   0.165&   5.774&    0.008&   0.00\\
 27&   0.6156805& 1&1&0&   4.553&   0.166&   7.395&    0.008&   0.00\\
 32&   0.6204257& 1&1&1& -30.317&  -1.107& -48.864&    0.006&   0.00\\
 33&   0.6023230& 1&1&1&   6.547&   0.239&  10.870&    2.714&   0.41\\
 34&   0.7339435& 1&1&1&   4.611&   0.168&   6.283&    0.005&   0.00\\
 38&   0.7790480& 1&1&1&   3.636&   0.133&   4.667&    0.307&   0.08\\
 40&   0.6340938& 1&1&1&   2.128&   0.078&   3.356&    0.007&   0.00\\
 41&   0.6629500& 1&1&1&  -3.317&  -0.121&  -5.003&    1.517&   0.46\\
 44&   0.5675425& 1&1&0&  -1.144&  -0.042&  -2.016&    0.373&   0.33\\
 45&   0.5891330& 0&1&1&  -1.325&  -0.048&  -2.249&    2.560&   1.93\\
 46&   0.6869440& 1&1&1&   3.922&   0.143&   5.710&    1.278&   0.33\\
 49&   0.6046400& 1&1&1&  -3.881&  -0.142&  -6.418&    6.582&   1.70\\
 51&   0.5741345& 1&1&1&   5.128&   0.187&   8.931&    0.616&   0.12\\
 54&   0.7729000& 1&1&1&   5.064&   0.185&   6.552&    0.008&   0.00\\
 55&   0.5816930& 1&1& & -13.428&  -0.490& -23.085&    0.008&   0.00\\
$^*56$&0.5680000& 0&0& &  23.585&   0.861&  41.523&   15.562&   0.66\\
 57&   0.7944180& 1&1&1&   4.800&   0.175&   6.042&    0.000&   0.00\\
 59&   0.5185200& 0&0& &  24.242&   0.885&  46.752&    5.406&   0.22\\
 62&   0.6197940& 1&1&1&   4.833&   0.176&   7.798&    1.194&   0.25\\
 63&   0.8259450& 1&1& &   6.706&   0.245&   8.119&    0.004&   0.00\\
 67&   0.5644508& 0&1& &   0.000&   0.000&   0.000&    0.000&   0.00\\
 69&   0.6532205& 0&1& &   4.463&   0.163&   6.832&    0.070&   0.02\\
 73&   0.5752130& 0&1& &  -5.269&  -0.192&  -9.160&    2.941&   0.56\\
 74&   0.5032475& 1&1&1& -10.883&  -0.397& -21.626&    3.649&   0.34\\
 79&   0.6082747& 1&1&1&   2.992&   0.109&   4.918&    0.426&   0.14\\
$^*84$&0.5798730& 1&1&0&  -1.375&  -0.050&  -2.371&    0.007&   0.01\\
 85&   0.7427580& 1&1&1&   2.461&   0.090&   3.313&    0.008&   0.00\\
 86&   0.6478330& 1&1&1&   3.239&   0.118&   5.000&    3.635&   1.12\\
 88&   0.6901980& 0&1& &  10.756&   0.393&  15.583&    2.290&   0.21\\
 90&   0.6034000& 1&1&1&   0.683&   0.025&   1.132&    0.006&   0.01\\
 91&   0.8951200& 1&-& &  47.375&   1.729&  52.926&    0.011&   0.00\\
 96&   0.6245312& 1&1& &   0.636&   0.023&   1.019&    0.008&   0.01\\
 97&   0.6918880& 1&1&1&   0.000&   0.000&   0.000&    0.000&   0.00\\
 99&   0.7661020& 1&1&2&  45.869&   1.674&  59.874&    2.248&   0.05\\
100&   0.5527119& 1&1& &  20.543&   0.750&  37.167&    6.698&   0.33\\
102&   0.6913905& 1&1&1&   4.611&   0.168&   6.669&    0.704&   0.15\\
104&   0.8676000& 1&1&0&-542.347& -19.796&-625.112&    1.460&   0.00\\
106&   0.5699070& 1&-& &   0.000&   0.000&   0.000&    0.000&   0.00\\
107&   0.5141010& 1&1&0&   1.811&   0.066&   3.523&    0.005&   0.00\\
108&   0.5944533& 1&1&1&   3.061&   0.112&   5.149&    0.708&   0.23\\
109&   0.7440700& 1&1&1&  13.528&   0.494&  18.181&    0.007&   0.00\\
111&   0.7629000& 1&1&1&   2.306&   0.084&   3.022&    0.004&   0.00\\
112&   0.4743565& 0&1&0&   0.389&   0.014&   0.820&    0.005&   0.01\\
113&   0.5733640& 1&1&1&   5.794&   0.211&  10.106&    1.624&   0.28\\
115&   0.6304640& 0&1&1&  -3.856&  -0.141&  -6.115&    2.456&   0.64\\
116&   0.7201320& 1&1& &  -1.989&  -0.073&  -2.762&    0.005&   0.00\\
118&   0.6116235& 1&1&1&  -3.028&  -0.111&  -4.950&    0.007&   0.00\\
120&   0.5485700& 0&1& & -11.406&  -0.416& -20.791&    1.067&   0.09\\
122&   0.6349230& 1&1&1&   1.944&   0.071&   3.062&    1.466&   0.75\\
125&   0.5928870& 1&1&1&  -0.353&  -0.013&  -0.595&    1.459&   4.14\\
128&   0.8349825& 1&1&1&   8.825&   0.322&  10.569&    0.838&   0.09\\
130&   0.4932500& 0&1& &   0.000&   0.000&   0.000&    0.000&   0.00\\
132&   0.6556390& 1&-& &   0.000&   0.000&   0.000&    0.000&   0.00\\
134&   0.6529060& 1&-& &  -4.762&  -0.174&  -7.294&    5.980&   1.26\\
139&   0.6768700& 1&-&1&   6.662&   0.243&   9.843&    0.010&   0.00\\
144&   0.8353110& 1&-& &   2.729&   0.100&   3.267&    0.009&   0.00\\
149&   0.6827280& 1&-& &  -5.069&  -0.185&  -7.425&    0.008&   0.00\\
150&   0.8992050& 1&-& &  67.825&   2.476&  75.428&   18.136&   0.27\\
\sidehead{RRc stars} 
010&   0.3749500& 0&0&&  14.676&   0.536&  39.140&   56.964&   3.88\\
012&   0.3867470& 1&1&&   4.436&   0.162&  11.470&    1.959&   0.44\\
014&   0.3771650& 1&1&&  18.650&   0.681&  49.448&   15.467&   0.83\\
016&   0.3301820& 1&1&&   7.803&   0.285&  23.632&    1.069&   0.14\\
019&   0.2995527& 1&1&&  -0.289&  -0.011&  -0.964&    0.103&   0.36\\
021&   0.3808160& 1&1&&  -4.892&  -0.179& -12.845&    7.146&   1.46\\
022&   0.3960800& 1&0&& -21.406&  -0.781& -54.044&   15.699&   0.73\\
024&   0.4622140& 1&1&&  11.592&   0.423&  25.079&    3.181&   0.27\\
030&   0.4041000& 0&0&& -21.397&  -0.781& -52.950&   28.643&   1.34\\
035&   0.3868382& 1&1&&  -2.800&  -0.102&  -7.238&    2.454&   0.88\\
036&   0.3798530& 1&0&&  -2.786&  -0.102&  -7.335&    6.196&   2.22\\
039&   0.3933700& 1&1&&   0.925&   0.034&   2.351&    3.223&   3.48\\
047&   0.4850500& 1&0&&  29.633&   1.082&  61.093&   62.330&   2.10\\
050&   0.3861840& 1&0&&  -8.892&  -0.325& -23.024&    1.847&   0.21\\
058&   0.3699170& 0&1&&   6.689&   0.244&  18.082&    1.876&   0.28\\
064&   0.3444720& 1&0&&  13.175&   0.481&  38.247&   21.644&   1.64\\
066&   0.4075400& 0&0&&  49.064&   1.791& 120.390&    5.776&   0.12\\
068&   0.5345500& 1&0&&  52.550&   1.918&  98.307&  100.053&   1.90\\
070&   0.3906000& 0&0&&   2.731&   0.100&   6.991&   21.733&   7.96\\
071&   0.3574650& 1&0&&   4.792&   0.175&  13.405&    4.314&   0.90\\
072&   0.3845250& 1&1&&  -6.979&  -0.255& -18.149&    9.579&   1.37\\
075&   0.4223000& 1&0&& -40.594&  -1.482& -96.127&   67.522&   1.66\\
076&   0.3378900& 1&0&&  37.025&   1.351& 109.577&   14.624&   0.39\\
077&   0.4260050& 0&0&&  12.778&   0.466&  29.994&    0.007&   0.00\\
081&   0.3893990& 1&1&&  -8.722&  -0.318& -22.399&    4.415&   0.51\\
082&   0.3358400& 1&0&& -45.956&  -1.677&-136.838&   10.129&   0.22\\
083&   0.3566080& 1&1&&   1.106&   0.040&   3.100&    0.004&   0.00\\
087&   0.3965100& 1&1&&  15.742&   0.575&  39.701&    1.416&   0.09\\
089&   0.3748533& 1&0&&  34.606&   1.263&  92.318&    6.808&   0.20\\
094&   0.2539330& 1&1&&   0.000&   0.000&   0.000&    0.000&   0.00\\
095&   0.4049850& 1&0&&   3.716&   0.136&   9.176&   23.817&   6.41\\
098&   0.2805655& 1&1&&   0.153&   0.006&   0.545&    0.008&   0.05\\
101&   0.3409200& 1&1&&  51.033&   1.863& 149.693&   14.825&   0.29\\
103&   0.3288490& 1&1&&   0.632&   0.023&   1.922&    0.009&   0.01\\
105&   0.3353340& 1&1&&   0.692&   0.025&   2.063&    0.435&   0.63\\
117&   0.4216600& 0&0&&   0.483&   0.018&   1.146&    6.754&  13.97\\
119&   0.3058765& 1&1&&  -1.639&  -0.060&  -5.358&    0.008&   0.00\\
121&   0.3041800& 1&0&&   1.981&   0.072&   6.511&    2.175&   1.10\\
123&   0.4739000& 0&0&& 149.964&   5.474& 316.446&  104.375&   0.70\\
124&   0.3318600& 1&1&&  -1.739&  -0.063&  -5.240&    0.332&   0.19\\
126&   0.3419100& 1&0&&  15.950&   0.582&  46.650&    8.616&   0.54\\
127&   0.3052760& 1&1&&  -2.239&  -0.082&  -7.334&    0.883&   0.39\\
131&   0.3921200& 1&0&& -18.122&  -0.661& -46.216&    4.627&   0.26\\
151&   0.4078000& 1&-&& 145.770&   5.321& 357.455&   77.384&   0.53\\
155&   0.4139230& 1&1&&   7.162&   0.261&  17.304&    0.010&   0.00\\
160&   0.3973400& 1&0&&-278.918& -10.180&-701.962&  267.563&   0.96\\
163&   0.3132300& 1&-&&   0.000&   0.000&   0.000&    0.000&   0.00\\
$^*168$&0.3212950&1&-&&  -3.547&  -0.129& -11.041&    0.011&   0.00\\
\sidehead{RR Lyrae like stars} 
 43&   1.1568240& 1&1&&  15.383&  0.561&  13.297&  3.775& 0.25 \\
 48&   4.4741000& 1&1&& 422.683& 15.428&  94.473&  0.000& 0.00 \\
 52&   0.6603750& 1&0&&   3.978&  0.145&   6.024&  1.071& 0.27 \\
 60&   1.3494300& 1&1&& 132.858&  4.849&  98.455& 24.940& 0.19 \\
 61&   2.2735550& 1&1&&  30.947&  1.130&  13.612&  4.204& 0.14 \\
 92&   1.3452000& 1&0&& 381.947& 13.941& 283.933& 14.064& 0.04 \\
\enddata
\tablenotetext{a}{ *: not radial velocity members \citep{li81,le00}.}
\tablenotetext{b}{ 0: light-curve showing any type of modulation.}
\tablenotetext{c}{ 0: uncertain \oc fit;  
--: not enough data to draw firm conclusion about long term period changes.}
\tablenotetext{d}{ 1: [Fe/H]$=-1.54\pm0.08$ dex; 0: less metal poor; 2: more metal poor stars 
(see Table~\ref{paramab}).}
\end{deluxetable}

\begin{deluxetable}{rrcccc}
\tablecolumns{6}
\tablewidth{0pt}
\tablenum{7}
\tabletypesize{\scriptsize}
\tablecaption{Physical parameters of RRab stars determined from their light-curves.\label{paramab}}
\tablehead{
\colhead{Var.}
&\colhead{$\beta$ ($d/10^{10}d$)} 
&\colhead{[Fe/H]}
&\colhead{Log$T_{eff}$} &\colhead{Log$L/L_{\sun}$} &\colhead{Log$M/M_{\sun}$}
}
\startdata
\sidehead{stars with [Fe/H]$=1.54\pm0.08$ dex}
3	&21.75	&-1.41	&3.8040	&1.822	&0.62\\
4	&3.73	&-1.60	&3.8192	&1.751	&0.64\\
7	&6.83	&-1.64	&3.8104	&1.783	&0.65\\
8	&16.95	&-1.51	&3.8304	&1.713	&0.67\\
13	& irr.	&-1.59	&3.8151	&1.760	&0.63\\
15	&20.41	&-1.54	&3.8030	&1.811	&0.64\\
18	& irr.	&-1.57	&3.8208	&1.750	&0.64\\
20	& irr.	&-1.52	&3.8212	&1.744	&0.64\\
25	& irr.	&-1.63	&3.8227	&1.739	&0.66\\
26	&4.53	&-1.53	&3.8037	&1.792	&0.63\\
32	&-30.32	&-1.56	&3.8211	&1.751	&0.64\\
33	&6.55	&-1.55	&3.8226	&1.741	&0.64\\
34	&4.61	&-1.62	&3.8077	&1.784	&0.64\\
38	&3.64	&-1.37	&3.8033	&1.781	&0.62\\
40	&2.13	&-1.60	&3.8187	&1.755	&0.65\\
41	&-3.32	&-1.51	&3.8163	&1.756	&0.62\\
45	& irr.	&-1.55	&3.8246	&1.738	&0.64\\
46	&3.92	&-1.53	&3.8133	&1.766	&0.63\\
49	& irr.	&-1.63	&3.8195	&1.737	&0.65\\
51	&5.13	&-1.52	&3.8257	&1.731	&0.65\\
54	&5.06	&-1.38	&3.8064	&1.781	&0.60\\
57	&4.8	&-1.46	&3.8030	&1.790	&0.62\\
62	&4.83	&-1.57	&3.8210	&1.749	&0.64\\
74	&-10.88	&-1.45	&3.8320	&1.704	&0.68\\
79	&2.99	&-1.41	&3.8230	&1.734	&0.62\\
85	&2.46	&-1.58	&3.8073	&1.784	&0.63\\
86	& irr.	&-1.57	&3.8170	&1.755	&0.64\\
90	&0.68	&-1.45	&3.8247	&1.737	&0.62\\
97	&0.00	&-1.65	&3.8112	&1.775	&0.66\\
102	&4.61	&-1.63	&3.8119	&1.773	&0.65\\
108	&3.06	&-1.58	&3.8227	&1.737	&0.65\\
109	&13.53	&-1.54	&3.8103	&1.795	&0.63\\
111	&2.31	&-1.33	&3.8081	&1.772	&0.59\\
113	&5.79	&-1.51	&3.8263	&1.732	&0.65\\
115	& irr.	&-1.67	&3.8191	&1.760	&0.66\\
118	&-3.03	&-1.64	&3.8195	&1.741	&0.65\\
122	& irr.	&-1.60	&3.8184	&1.754	&0.65\\
125	& irr.	&-1.49	&3.8248	&1.736	&0.63\\
128	&8.83	&-1.46	&3.7998	&1.811	&0.63\\
139	&$-$	&-1.58	&3.8135	&1.756	&0.63\\
\sidehead{less metal poor stars}
23	& irr.	&-1.07	&3.8326	&1.666	&0.60\\
27	&4.55	&-0.80	&3.8214	&1.677	&0.54\\
44	&-1.14	&-1.16	&3.8261	&1.695	&0.60\\
84	&-1.38	&-1.01	&3.8227	&1.677	&0.57\\
104	&-542.35&-0.74	&3.7996	&1.773	&0.55\\
107	&1.81	&-1.13	&3.8338	&1.682	&0.61\\
112	&0.39	&-1.19	&3.8347	&1.663	&0.65\\
\sidehead{more metal poor star}
99	&45.87	&-1.99	&3.8070	&1.840	&0.71\\
\enddata

\tablecomments{The physical parameters are calculated using the formulae given in
\citet{j98a}, the Log$L$ and Log$T$ scales are shifted by 0.1 and 0.016, respectively,
in order to reach agreement with the \citet{do92} evolutionary models \citep{jk99}.}
\end{deluxetable}

\end{document}